\documentstyle[aps,epsfig]{revtex}

\begin{document}

\twocolumn[\hsize\textwidth\columnwidth\hsize\csname 
@twocolumnfalse\endcsname
\title{Effect of magnetic field on over-doped HTc superconductors: 
Conflicting predictions of various HTc theories}
\author{Oron Zachar}
\address{Laboratoire de Physique des Solides, U. Paris-Sud, 91405 Orsay , France.}
\maketitle

\widetext
\begin{abstract}
Following the recent $NMR$ experiments by Gorny et al., we discuss the effect of 
a magnetic field on the superconducting $T_{c}$ and the spin pseudo-gap 
$T^{*}$. As a testable prediction, we argue that a spin pseudo-gap should also 
be observed in over-doped samples in a magnetic fied (while normally there is 
no pseudo-gap above $T_{c}$). We find that different theoretical approaches have 
marked differences in their predictions for over-doped HTc cuprates. 
\medskip 
\end{abstract}

 ]

\narrowtext

\section{Introduction}

In a high accuracy $NMR$ experiment on near optimal doped $YBCO$, Gorny et
al.\cite{Gorny98-NMR-pseudogap} found that a magnetic field of $14.8$ Tesla
shift $T_{c}$ down by as much as $8K$, while the spin pseudo-gap remains
unaffected (as measured by $\left( T_{1}T\right) ^{-1}$). They concluded
that it is an evidence that ''...hence the pseudo-gap is unrelated to
superconducting fluctuations''.

Here, we make a {\sl testable prediction}: That, {\sl over-doped} $HT_{c}$
samples (which normally do not show a spin pseudo-gap state) when subjected
to a moderate magnetic field will unprecedentedly reveal a spin pseudogap
above $T_{c}\left( B\right) $ starting from approximately the original 
$T_{c}\left( B=0\right) $ of zero magnetic field\cite{note1-Gorny-new}. Our
prediction is based both on a new phenomenological interpretation of the
experiments (in contrast with the original interpretation of Gorny et al.),
and further on a microscopic stripes theoretical approach. 

As elaborated below, the above prediction is {\sl not} shared by several
other current HTc theories. The general difference in the prediction of
various theoretical approaches for the effect of a magnetic field on
over-doped HTc cuprates can be understood from the difference in the
position of the spin pseudo-gap line, $T^{\ast }$, in the two theoretical
phase diagrams depicted in {\sl Figure-1}. Therefore, repeating the
experiments of Gorny et al. \cite{Gorny98-NMR-pseudogap} on overdoped HTc
samples constitute a crucial experiment to determine the proper form of the
HTc phase diagram in the over-doped region, and hence to provide further
theoretical constraints.

\section{Two phase diagrams: Conflicting predictions}

The spin pseudo-gap is often referred to as a peculiarity of underdoped and
near optimal doped HTc superconductors, where the gap evolves smoothly as
the temperature is increased through $T_{c}$ and remains significant up to a
cross-over temperature $T^{*}>T_{c}$. This is in sharp contrast to the
behavior of over-doped cuprates (doping $x>0.2$) and conventional
superconductors where the gap closes at $T\geq T_{c}$. Hence, it is common
to find references to the over-doped cuprates as more conventional, (even by
theorists who otherwise advocate non-conventional mechanisms).

\begin{figure}
\begin{center}
\leavevmode\epsfxsize=2.5in 
\epsfbox{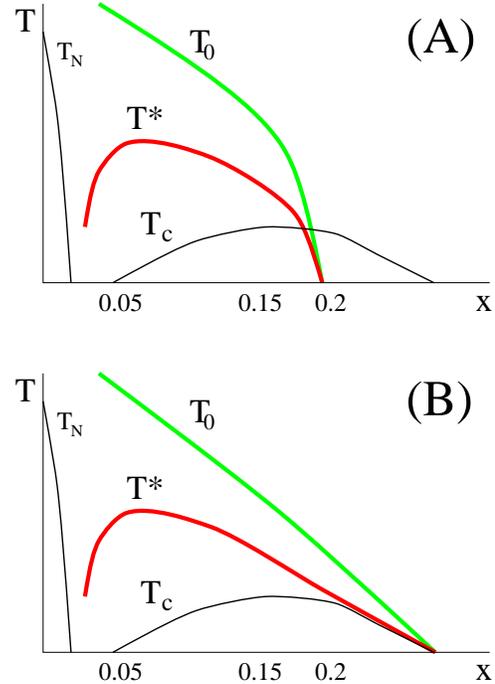} 
\end{center}
\caption{Theoretical phase diagrams}
\label{HTc-PhaseDiagram-fig}
\end{figure}

At the under-doped and optimal regions of doping ($0.05<x<0.2$), there is a
growing agreement that two cross-over temperatures can be identified; A
doping dependent cross-over temperature $T_{0}\left( x\right) $ is
experimentally marked by a broad maxima in the spin susceptibility $\chi
_{0}\left( T\right) $. In addition, below $T_{0}\left( x\right) $, $T_{1}T$
decreases linearly in temperature and there is a downward deviation of the
in-plane resistivity $\rho _{ab}\left( T\right) $. At a lower temperature 
$T^{\ast }\left( x\right) $, a second cross-over occurs, when a pseudo-gap
feature appears in NMR, ARPES, neutron scattering, and specific heat
measurements. Below $T^{\ast }\left( x\right) $, $\chi _{0}\left( T\right) $
continues to decrease even more rapidly down to $T_{c}$, but $T_{1}T$ 
$\ $
exhibit a minimum followed an increase as temperature is lowered further,
which is suggestive of a spin gap formation. (sometime both cross-overs are
referred to as ''pseudo-gaps'', which remains a cause for confusion).

In the over-doped region, the continuation of the pseudo-gap lines below 
$T_{c}$ in Figure-1 should be understood as ''what would be if there was no
superconducting phase'', which is exactly what a magnetic field does\cite
{Boebinger96}. The difference between phase diagrams (A) and (B) in Figure-1
is in the over-doped region, and the proposed NMR experiment will reveal the
correct one\cite{note2} (and thus pose a challenge to the other theories).

{\bf (a)} Figure-1A is the one most commonly found in the literature\cite
{note2}. Near an over-doping point $x_{over}\approx 0.2$, where $T^{\ast
}\approx T_{0}=T_{c}$, the pseudo-gap lines end sharply (i.e., cross the 
$T_{c}$ line). Explicit examples of such phase diagrams are currently drown
by Pines and collaborators \cite{Pines98-HTc-review} (which advocate a spin
fluctuation exchange mechanism), and by the Rome group of Castellani,
DiCastro, Grilli and collaborators\cite{Castellani98-HTc-review} (which
advocate a quantum critical point fluctuations mechanism). In addition, if
the pseudo-gap below $T^{\ast }$ is not related to pairing then there is no
reason for it to be correlated with $T_{c}$ in the over-doped region. The
choice of the $x_{over}\approx 0.2$ point is not arbitrary. There are
various experimental indicators for a significant qualitative change in the
cuprates beyond this point; A prime example is the experiment of Boebinger
\cite{Boebinger96} which implies a metal-insulator quantum phase transition.
The physical significance of each cross-over line (and the $x_{over}\approx
0.2$ point) is, of course, varying between theories.

For over-doping $x>x_{over}\approx 0.2$ in Figure-1A, it means that $T^{\ast
}\approx T_{0}<T_{c}$. Therefore, if Figure-1A is the correct phase diagram
then for an over-doped HTc sample under a magnetic field $B$ of about $8-14$ 
$Tesla$, the following predictions are implied:

\begin{enumerate}
\item  Though $T_{c}$ will be suppressed by a few degrees, there will remain
no signature of a spin pseudo-gap behavior above $T_{c}\left( B\right) $.

\item  In particular, $\chi _{0}\left( T\right) $ will continue to {\sl %
increases} with decreasing temperature down to $T_{c}$, in sharp contrast
with a pseudo-gap behavior where below $T^{\ast }$ it {\sl decreases}
rapidly down to $T_{c}$.
\end{enumerate}

{\bf (b)} In contrast, we now introduce arguments in favor of the phase
diagram depicted in Figure-1B, where the $T^{\ast }$ cross-over line merges
continuously with $T_{c}$, but it is still ''there'' (as a pairing
mechanism) and can be revealed in the appropriate NMR experiment.

To explain the pseudo-gap phenomenon below $T^{\ast }$, a general argument
base on superconducting phase fluctuations was introduced by Emery\&Kivelson
\cite{EK-PhaseFluctuations} and elaborated by Millis and collaborators \cite
{Millis98-PhaseFluctuations}. As depicted in figure-2, the superconducting
transition temperature $T_{c}$ is determined by the lowest of two
parameters; the pairing temperature $T_{pair}\sim \Delta \left( 0\right) /2$
, and the phase ordering temperature $T_{\theta }$. The establishment of a
significant pairing amplitude is determined by the pairing energy scale
which is given by the spin gap $\Delta \left( T\right) $. The classical
phase ordering temperature $T_{\theta }$ is obtained by considering the
disordering effects of only the classical phase fluctuations as $T_{\theta
}\sim V_{0}$, where $V_{0}={\frac{\hbar ^{2}n_{s}(0)a}{4m^{\ast }}}$ is the
zero-temperature value of the ``phase stiffness'' (which sets the energy
scale for the spatial variation of the superconducting phase). In
conventional weak coupling BCS superconductors $T_{\theta }\gg T_{pair}$ and
hence $T_{c}=T_{pair}$.

Note that the phase stiffness can be reduced either by increasing the
effective quasiparticle mass (i.e., when $m^{\ast }\gg m_{e}$), or by low
superfluid density $n_{s}(0)$. It is argued that in the HTc cuprates 
$n_{s}(0)$ is indeed low enough that phase fluctuations become important (see
further discussion below). The density of mobile charge carriers, and hence
the superfluid density $n_{s}(0)$, naturally increases with increased doping 
$x$.

\begin{figure}
\begin{center}
\leavevmode\epsfxsize=2.5in 
\epsfbox{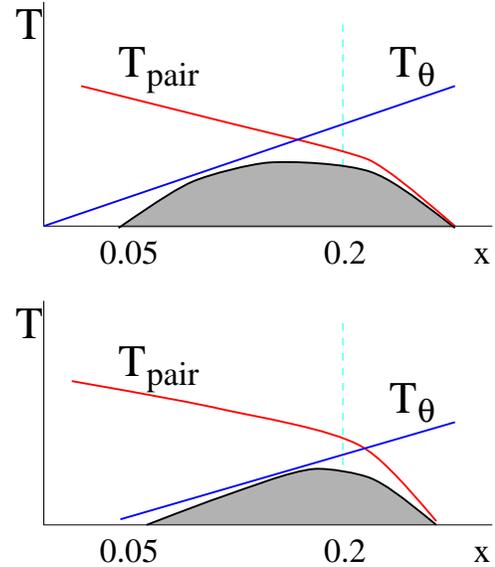} 
\end{center}
\caption{The upper figure depicts the usual plot, in zero magnetic field, 
showing that $T_c$ is limited by $T_{\theta }$ in the case of low 
superfluid density $n_{s}(0)~x$, and by $T_{pair}$ at higher $n_{s}(0)$. 
A pseudo-gap correspond to the region where $T_{\theta } < T_{pair}$ . The dashed 
line at the putative $x=0.2$ doping serves as a guide to the eye. The lower 
figure depicts our suggested effect of a magnetic field for the case of 
HTc superconductors. As explained in the text, the $T_{\theta }$ line is shifted 
down while $T_{pair}$ remains nearly unsuppressed. As a result, the pseudo-gap 
region now extends to higher dopings.}
\label{HTc-PhaseFluctuation-fig}
\end{figure}

Figure-2A depicts the resulting theoretical phase diagram in the absence of
a magnetic field\cite{EK-PhaseFluctuations} (and neglecting competition with
other order parameters such as AFM). In the under-doped and optimal-doped
regions, where $T_{pair}>T_{\theta }$, $T_{c}$ is determined by the phase
ordering temperature $T_{\theta }$. In particular, there is a temperature
range $T_{c}<T<T_{pair}$ where there is significant pairing amplitude
without global phase coherence. Therefore, within this framework we make the
identification of the spin pseudo-gap temperature $T^{\ast }=T_{pair}$. In
contrast, in the over-doped region, $T_{c}$ is determined by $T_{pair}$,
i.e., by the pairing energy scale as in conventional weak coupling BCS.

Focussing here on the over-doped region, one would be first lead to the
conclusion that the above picture entails that the effect of a magnetic
field on an over-doped sample would be similar to the case of conventional
superconductors.

It is important to understand that the microscopic pairing mechanism and the
phase coherence mechanism are distinct. The effect of a magnetic field is
sensitive to microscopic details which are not part of the macroscopic phase
fluctuation theory. In weak coupling BCS theory, a significant part of the
pairing energy is an outcome to the overlap between the pairs (due to the
large pair coherence length), which leads to coherent scattering among many
pairs within the single pair coherence length. In this sense, phase
coherence (on a local scale) between pairs is affecting also the pairing
scale (i.e., the gap magnitude) even though phonons are not affected. This
is not the case in HTc superconductors where the $d-wave$ pair coherence
length is on the order of one lattice spacing.

Interpreting the experimental findings of Gorny et al.\cite
{Gorny98-NMR-pseudogap} within the above framework, we are led to the
statement that {\sl a magnetic field suppresses the superconducting phase
coherence, while the pairing mechanism (leading to the spin pseudo-gap)
remains much less affected}. (At present we draw this conclusion
phenomenologically, irrespective of microscopic origins. A candidate
microscopic theory leading to such phenomena will be discussed below). Thus,
the effect of a moderate magnetic field on the phase diagram is as depicted
in {\sl Figure-2B}. The down shift of $T_{\theta }$ by a magnetic field
requires some explanation: In the absence of a magnetic field, the phase
coherence temperature $T_{\theta }$ is a result only of phase fluctuations.
Within Ginzburg-Landau theory, the magnetic field energy density is
competing with the condensation energy, i.e., with phase coherence.
Therefore, in the presence of a magnetic field, $T_{\theta }$ is determined
by adding the phase fluctuation contribution on top of the usual phase
coherence frustrating effect of the magnetic field. For an over-doped HTc
sample in a magnetic field, the following predictions are implied:

\begin{enumerate}
\item  Under a sufficiently strong magnetic field $H$ (e.g., $8-14$ Tesla),
a spin pseudo-gap will be revealed at $T^{\ast }\approx T_{c}\left(
H=0\right) $ above $T_{c}\left( H\right) $ at approximately the original 
$T_{c}$ in the absence of a magnetic field. This is an unprecedented
prediction.

\item  The dependence of $T_{c}\left( H\right) $ on the magnetic field will
also be quite peculiar, and different from what is observed in under-doped
samples. Initially, when still $T_{\theta }\left( H\right) >T_{pair}\left(
H\right) =T_{c}\left( H\right) $, there will be very little suppression of 
$T_{c}\left( H\right) $ due to the relatively small suppression of the
pairing mechanism. Yet, below a critical magnetic field $T_{\theta }\left(
H_{cr}\right) =T_{pair}\left( H_{cr}\right) =T_{c}\left( H_{cr}\right) $,
there will be a more rapid suppression of $T_{c}$ with increased magnetic
field (since now $T_{c}\left( H\right) =T_{\theta }\left( H\right)
<T_{pair}\left( H\right) $).
\end{enumerate}

We add a remark that the effect of a magnetic field on phase coherence and
pair fluctuation effects may depend on microscopic details. 
In a weak coupling BCS model of a d-wave superconductor, Esching et al.\cite
{Sauls99-MagField-PairFluctuations} concluded that the a magnetic field
reduces also the fluctuation corrections to $\left( T_{1}T\right) ^{-1}$,
i.e., lead to also lower $T^{\ast }$. Similarly, Pines (\cite
{Pines98-HTc-review}, page:14) state that moderate magnetic fields will have
a dephasing effect on the pairing channel (via AFM spin fluctuation
exchange) and thus significantly suppress $T^{\ast }$ (in contrast with our
phenomenological assumption above, and with the stripe model described
below).

Emery-Kivelson and Collaborators\cite{SpinGapProximity} elaborated a
theoretical approach to HTc which is based on coupled fluctuating spin and
charge stripes in real space. The stripes are local phase separated
electronic structures made out of quasi one-dimensional hole rich conducting
electronic filaments (referred to as ''hole-lines'') and confined in between
them are narrow ladder-like half-filled regions (which thus have substantial
AFM correlations).

The systematics of phase fluctuations\cite{EK-PhaseFluctuations}, mentioned
above, suggests that pairing on a high energy scale does not require
interaction between metallic charge stripes. Instead, pairing is established
first on single stripes, independently, at temperature $T^{\ast }$ (the
single stripe is modelled as a 1D electron gas coupled to the various
low-energy states of an insulating ladder-like environment\cite
{SpinGapProximity}). Below $T^{\ast }$, each charge stripe can be regarded
as a spin gaped one dimensional extended ''grain'' with enhanced pairing. In
turn, $T_{c}$ is controlled by the Josephson coupling required to establish
phase coherence for an array of stripes\cite{SpinGapProximity}.


Another way of looking at the situation is to compare the superfluid density 
$n_{s}(0)$ with the number of particles $n_{P}$ involved in pairing. In BCS
theory, at $T=0$, $n_{P}$ is of order $\Delta _{0}/E_{F}$ (where E$_{F}$ is
the Fermi energy) and $n_{s}(0)$ is given by all the particles in the Fermi
sea; {\it i.e.} $n_{p}\ll n_{s}(0)$. For Bose condensation $n_{P}=n_{s}(0)$.
In the stripes model of {high temperature superconductors}, $n_{P}\gg
n_{s}(0)$; most of the electrons in the Fermi sea participate in the spin
gap below $T^{\ast }$ (since both the electronic ladder environment and the
hole-lines develop the spin gap) but the superfluid density of the doped
insulator is small, since the mobile charge density, proportional to $x$,
includes only the charges in the hole lines.

Though the stripes are extended objects, their effective one-dimensionality
entails that a magnetic field does not significantly alter the electron
pairing dynamics on individual stripes, and thus $T^{\ast }\left( H\right) $
is predicted to remains almost constant. Similarly, the lack of large
fluctuation diamagnetism between $T^{\ast }$ and $T_{c}$ is readily
understood, since an applied magnetic field does not drive any significant
orbital motion until coherence develops in two (and ultimately three)
dimensional patches, close to $T_{c}$\cite{SpinGapProximity}. Below $T^{\ast
}$, Josephson coupling between stripes leads to the establishment of global
phase coherence. Hence, as in conventional granular superconductors, a
magnetic suppresses the phase coherence between stripes.

{\sl In conclusion, the effect of a magnetic field on the microscopic
dynamics of the stripes model leads to the same predictions which where
deduced above following a phenomenological re-interpretation of the
experiments of Gorny et al due to the separation of pairing an phase
coherence scales}.

As an additional remark, notice that the under-doped end of the $T^{\ast }$
line is drawn as going down sharply towards zero below $x=0.04$. This is
also a testable consequence of the spin-gap-proximity effect mechanism in
the stripes approach\cite{SpinGapProximity}. Between doping $x=0.06$ and 
$x=0.02$, we may envision two extreme scenarios leading to the same
consequence for the spin gap of the effective AFM ladder environment
(between each two hole rich lines), which in turn affect the total spin gap:
1) If the hole lines filling fraction remains constant then the AFM ladder
triples its width, which implies that the theoretical maximum spin gap
decreases by a factor $e^{-3}\approx \frac{1}{20}$. 2) If the width of the
ladder environment remains constant then the hole-lines filling
''overflows'' and dissolves the stripe structure.

\section{Summary}

The NMR experiment of Gorny et al. indicate that a magnetic field shifts 
$T_{c}$ down while the spin pseudo-gap (what ever is its origin) remains
relatively unaffected (i.e., $T^{\ast }\sim $constant). We point out that in
any model in which superconducting pairing and phase ordering are governed
by distinct physics, that distinct dependences on parameters of the pairing
scale and the superconducting $T_{c}$ are to be expected, in conflict with
the conclusion of Gorny {\sl et al}\cite{Gorny98-NMR-pseudogap}. If the
pseudo-gap is associated with local pairing then it implies that a magnetic
field is only weakly suppressing the pairing energy scale (unlike weak
coupling BCS).

Current theoretical approaches where conceived to agree with the known
experimental results in under-doped and optimal-doped cuprates. Yet, their
implied characterization of the over-doped region ($x>0.2$) are distinct. As
depicted in Figure-1, the difference is highlighted by the continuation of
the $T^{\ast }$ line in the over-dope region. In this paper we argued that
NMR experiments can reveal those differences.

In particular, we present the following argument and prediction: (1) Let us
assume that there is only one and the same mechanism of HTc
superconductivity in all the cuprates and over the whole doping range (from
under to over doping\cite{note4}). (2) Assume that the spin pseudo-gap below 
$T^{\ast }$ is indeed a precursor of the superconducting gap, i.e., an
outcome of a developed local pairing amplitude in the absence of global
phase coherence. (3) The over-doped cuprates are characterized by $%
T_{c}=T^{\ast }$ in the absence of an external magnetic field. (4) It
follows from the experiment of Gorny et al.\cite{Gorny98-NMR-pseudogap} that 
{\sl a magnetic field suppresses the superconducting phase coherence
temperature }$T_{c}${\sl , while the pairing mechanism\ remains much less
affected (as measured by }$\left( T_{1}T\right) ^{-1}${\sl ). Therefore, we
predict that in an over-doped sample (}$x>0.2${\sl ) in a moderate magnetic
field (}$8-14${\sl \ Tesla) a spin pseudo-gap will be revealed starting from
approximately the original }$T_{c}$,{\sl \ at }$T^{\ast }\approx T_{c}\left(
H=0\right) $,{\sl \ above }$T_{c}\left( H\right) ${\sl \ }(while there was
no pseudo-gap in the absence of a magnetic field).

The above prediction is a natural consequence of the spin-gap-proximity
effect mechanism advanced by Emery-Kivelson-Zachar \cite{SpinGapProximity},
but is not shared by several other leading theoretical approaches\cite
{Pines98-HTc-review,Castellani98-HTc-review,note2}. Hence, the result of
performing the suggested $NMR$ experiment can serve to provide new
theoretical constraints; On the one hand, if the pseudo-gap phenomenon will
prove to be only a curiosity of underdoping then it does not reflect an
essential part of the pairing mechanism. On the other hand, if the above
suggested experiment will unprecedentedly reveal a pseudo-gap region also in
over-doped samples then it will strengthen the view that the underdoped
materials exemplify the essential physics of HTc, which is only getting
progressively obscured (due to similar energy scale of otherwise distinct
phenomena) in optimal and overdoped samples, and not vice versa.




\begin{references}
\bibitem{note2}  The over-doped region of the phase diagram is at many
instances not extensively elaborated by the original authors of each theory.
Hopefully this article will have a stimulation affect to change that. I made
reference only to theories which presented a phase diagram which was
elaborated enough to draw some conclusion from. Unfortunately, there aren't
many.
\bibitem{note3}  Some reservation needs to be added, that the magnetic field
should not be so strong as to significantly suppress also the
antiferromagnetic correlations which are related to the origin of the spin
pseudo-gap in many theoretical approaches.
\end{references}
\end{document}